# Advancing the Theory and Understanding of Quantum Self-Frictional Fields

B. A. Mamedov

*Department of Physics, Faculty of Arts and Sciences, Gaziosmanpasa University, Tokat, Turkey*

*e-mail: bamamedov@yahoo.com*

## Abstract

The self-friction field of electrons remains one of the fundamental problems in classical and quantum physics, retaining its significance and continuing to constitute a primary focus of both theoretical and experimental research. This study aims to discuss the recent important developments in the examination of the quantum self-frictional field, which is one of the most important old physics problems, and includes the research direction that needs to be examined in the future. In this study, we have presented a detailed explanation of the physical nature of quantum self-frictional atomic field potentials, including enhancements and developments on the significant subject. A complete understanding of classical and quantum self-frictional field potential and force has been carefully examined.

## I. Introduction

It is known that the leading important theoretical discoveries in physics that were consistent with the experimental observations have made essential contributions to the progress of the applied sciences and technology [1-10]. During the last century, considerable progress have been achieved in the experimental and theoretical studies of the physical nature of matter, for instance, Bohr's structure of atoms and the radiation emanating from them [1-6], the photoelectric effect [1-6], the Rutherford model of the atom [1-6], the Einstein's theory of special and general relativity [11-19], the mass-energy equivalence [20-24], the detection of gravitational waves [25-29], the quantum entanglement [30-45], the Higgs boson [46, 47], the observed Higgs boson [48-53], the classical Lorentz damping (self-frictional) theory [54-56], the laser [57-58], and others were first discovered theoretically and then proved experimentally. As can be seen from the literature, these kinds of theoretical discoveries in the humanities have taken their significant place in practice with very good experience throughout more than the last 50 years [1-4]. Note that in physics today there are phenomena that have been discovered theoretically but are expected to be observed experimentally, for example, we can highlight physics beyond the standard model, the Lorentz dumping theory, neutrino mass, matter-antimatter asymmetry, and others [51-56]. First, Lorentz theoretically explained that the electron could remain under the influence of its self-friction field. This study reviews some of this earlier knowledge of atomic self-friction field theory and also presents recent important developments in the field, as detailed in the sections. Lorentz proposed an approximation according to the classical electromagnetic theory of damping, in which electrons move around the nucleus under the action of nuclear and self-friction interaction forces [54-56]. We note that from time to time, based on the classical theory of the self-friction field proposed by Lorentz, an attempt has been made in the literature to explain the quantum theory of the self-friction field and to derive the related expressions [59-62]. Unfortunately, despite all the efforts in the quantum theory of the self-friction field, due to several reasons, successful results were not achieved. In recent years, there has been considerable progress in the theoretical investigation of the quantum nature of the electron self-friction field [63-67]. Guseinov also made important contributions to the progress of other fundamental problems of physics for future developments in science. His most notable contributions were as follows: generalized Dirac equation for particles of arbitrary mass and spin [77, 78], combined open-shell Hartree–Fock theory of atomic–molecular and nuclear systems [79], combined open-shell Hartree–Fock theory of atomic and molecular systems constructed from noncharged scalar particles with rest mass of particles $m=1$ and $m\neq 1$ [66, 80].

This study focuses on the contributions and applications of Guseinov and his collaborators in advancing the quantum theory of the classical self-friction field originally proposed by Lorentz in classical electrodynamics. Another objective of this work is to evaluate the reliability of these theoretical developments through practical applications. In particular, we analyze the overall theoretical progress to highlight the significance of the quantum nature of the atomic self-friction field and to outline future research directions.

## 2. Classical and quantum atomic self-friction field theory

The investigation of the physical reality of the atomic quantum self-friction field has recently attracted significant attention from the scientific community. According to the classical theory of self-friction proposed by Lorentz within the framework of classical electrodynamics [1–3], electrons move around atomic nuclei under the influence of both nuclear attraction and the field generated by their own damping (or self-gravitational) forces. Consequently, electrons are subject to two types of forces as they orbit the atomic nuclei: nuclear attraction and self-frictional forces, which are defined as follows [54–56]:

$$\vec{F} = \vec{F}_N + \vec{F}_L . \quad (1)$$

In Eq.(1), $\vec{F}_N$ is the nuclear attraction forces, $\vec{F}_L = \frac{2e^2}{3c^3}\dddot{\vec{r}}$ is the Lorentz self-frictional forces and $\dddot{\vec{r}}$ is the time derivative of the acceleration of the electron. The first successful quantum theory of the self-friction field of atomic electrons was developed by Guseinov. He proposed a general approximation framework based on quantum self-frictional nuclear attraction forces for calculating fully orthonormal one-electron $\psi^{(\alpha^*)}$ - self-frictional exponential-type wave functions ($\psi^{(\alpha^*)}$- SFETOs), which represent an extension of Lorentz's classical theory to quantum systems. These functions are formulated in the standard convention as follows [63–66]:

$$\psi_{nlm}^{(\alpha*)}\left(\zeta,\vec{r}\right)=\left(2\zeta\right)^{3/2} e^{-\frac{x}{2}} L_{nl}^{(\alpha*)}\left(x\right) S_{lm}\left(\theta,\varphi\right) \quad (2)$$

$$L_{nl}^{(\alpha*)}\left(x\right)=\left[\frac{\left(n\text{-}l\text{-}1\right)\,!}{\left(2n\right)^{\alpha^*}\Gamma\left(q*+1\right)}\right]^{1/2} x^l L_{n-l-1}^{(p*)}\left(x\right), \quad (3)$$

$$L_{n-l-1}^{(p*)}(x)=\frac{\Gamma(q^*+1)}{(n-l-1)!\,\Gamma(p^*+1)}\,{}_1F_1(-[n-l-1];p^*+1;x) \quad (4)$$

where, $x=2\zeta r$, $0<\zeta<\infty$, $q*=n+l+1-\alpha*$, $p*=2l+2-\alpha*$ and $\alpha*$ is the noninteger or integer ($-\infty<\alpha\leq 2$ for $\alpha*=\alpha$) frictional quantum number ($-\infty<\alpha*<3$ for $\alpha*\neq\alpha$). $S_{lm}\left(\theta,\varphi\right)$ are the complex or real spherical harmonics [81], $L_{n-l-1}^{(p*)}$ and $L_{nl}^{(\alpha*)}$ are the generalized and modified Laguerre polynomials, respectively. The ${}_1F_1\left(\eta;\gamma;x\right)$ is the confluent hypergeometric function and $\left(\eta\right)_k$ is the Pochhammer symbols, for definition see [82]. Note that, on the assumption of disappearing the self-friction forces (in the case of $\alpha*=1$ and $\zeta=Z/n$), the $\psi^{(\alpha*)}$-ETOs eigenvectors and eigenvalues are converting to Schrödinger's results in nonstandard convention for the hydrogen-like atom.

Based on the new consideration, presented in the study [63-66], a complete atomic quantum self-frictional potential is also developed in the following form:

$$V_{nl}^{(\alpha*)}(\zeta,r)=-\frac{\zeta n}{r}U_{nl}^{(\alpha*)}(x) \quad (5)$$

$$U_{nl}^{(\alpha^*)}(t)=1+\left(\frac{\alpha^*-1}{n}\right)\sqrt{2n(n-(l+1))}\,\frac{L_{nl+1}^{(\alpha^*+1)}(x)}{xL_{nl}^{(\alpha^*)}(x)}, \tag{6}$$

It is important to note that the self-frictional properties disappear for $\alpha^*=\alpha=1$, i.e., electron move only under the core nuclear attraction field:

$$V_{nl}^{(\alpha^*)}(\zeta,r)=-\frac{Z}{r} \tag{7}$$

Scientific analysis shows that one of the major difficulties in advancing the quantum theory of the Lorentz damping field force is its dependence on the time derivative of the electron's acceleration. Guseinov first proposed the quantum self-friction theory in physics by extending Lorentz's damping theory to assume that electrons move around atomic nuclei under two types of forces, nuclear attraction and self-friction. The main source of these forces is the quantum self-friction fields, corresponding to the classical damping fields introduced by Lorentz. We note that the quantum self-frictional theory in the standard convention is one of the great progresses in the investigation of multielectron atoms and molecules structure with the new perspective.

**3. Conclusion**

To confirm the validity of recent advances in the quantum theory of the self-friction field, this framework has been successfully applied to various systems, including small-sized atomic structures [67–76]. As an application, the dependence of quantum self-frictional potentials on the distance from the nucleus has been investigated using various $n$, $l$, $\alpha^*$ and $\zeta$ parameter values. It is clearly seen from the graphics that according to the variation of the position vector, the electron's full self-friction field potential have the attractive and repulsive properties. In the study [66], authors firstly present the approach for evaluating the average values of potentials, kinetic and total energies, and forces of the hydrogen-like atomic systems using self-friction complete orthonormal $\psi^{(\alpha^*)}$ - SFETOs with the various values of self-friction quantum numbers. The new development has been applied to the investigation of some closed and open shells atomic and molecular systems by using the Combined Hartree-Fock-Roothaan equations in the basis function $\psi^{(\alpha^*)}$ - SFETOs and variations of $\alpha$ SF quantum number and $\zeta$ scaling parameters [68-76]. The orbital, kinetic, and total energies and linear combination coefficients of some atoms and molecules have been calculated and analyzed for various values of $\alpha$ SF quantum number and $\zeta$ screening parameters. As can be seen from the data obtained, we can obtain more realistic and substantial results, for example, when we consider the possible interactions between the particles that make up the atomic and molecular systems.

The emerging progress in the quantum theory of the atomic self-friction field, one of the major developments in theoretical physics over the last 20 years, is expected to attract the attention of experimental physicists and be observed experimentally. We believe that the experimental observation of this phenomenon will lead to new and important results in physics, particularly concerning the electronic structure of atoms and molecules.

**Conflict of interest statement**

On behalf of all authors, the corresponding author states that there is no conflict of interest.

**References**

1. Murugeshan, R., Kiruthiga, S, Modern Physics, New Delhi, 2016.
2. Noce, C., Modern Physics: A Critical Approach, IOP Publishing, 2020
3. Weinberg, S., Foundations of Modern Physics, Cambridge Universite Press, 2021.
4. Singh, R.B., Introduction to Modern Physics, New Age International Publishers, New Delhi, 2009.
5. Burns, M.L., Modern Physıcs For Scıence And Engıneerıng, Physics Curriculum & Instruction, 2012.
6. Griffiths, D.J., Introduction to Quantum Mechanics, Cambridge University, 2017.
7. Zettili, N., Quantum Mechanics: Concepts and Applications, Wiley, UK, 2009.
8. Atkins, P.W., Friedman, R.S.. Molecular Quantum Mechanics, Oxford, 1996.
9. Greiner, W., Quantum Mechanics, Springer, New York, 2001.
10. Gluck, P., Agmon, D., Classical And Relativistic Mechanics,Word Scientific, London, 2009.
11. Kenyon, I. R., General relativity, Oxford University Press, Oxford (UK), 1990.
12. Foster, J., Nightingale, J. D., & Foster, J., A short course in General Relativity, New York: Springer-Verlag, 1995.
13. Soffel, M. H., Han, W. B. (2019). Applied general relativity. *Astronomy and Astrophysics Library, 2019*.
14. Stephani, H.. Relativity: An introduction to special and general relativity. Cambridge University Press, 2014.
15. Walecka, J. D., Introduction to general relativity. World Scientific Publishing Company, 2007.
16. Griffiths, J. B., Podolský, J., Exact space-times in Einstein's general relativity. Cambridge University Press, 2009.
17. Ziefle, R. G., Failure of Einstein's theory of relativity. I. Refutation of the theory of special and general relativity by an empirical experiment and by an epistemological analysis. Phys. Essays, *32*(2) (2019) 216-227.
18. Berti, E., Barausse, E., Cardoso, V., Gualtieri, L., Pani, P., Sperhake, U., ... & Zilhao, M., Testing general relativity with present and future astrophysical observations. Class. Quant. Gravity, *32*(24), (2015) 243001.
19. Brown, H. R. (1997). On the role of special relativity in general relativity. Int. Stud. Philos. Sci., 11(1) (1997) 67-81.

20. Fernflores, F., Einstein's Mass-Energy Equation, Volume 1, Momentum Press, 2017.
21. Francisco, F., The equivalence of mass and energy, 2001.
22. Okun, L. B., The concept of mass. Physics Today, *42*(6) (1989). 31-36.
23. Chang, D. C. A quantum interpretation of the physical basis of mass–energy equivalence, *Mod. Phys. Lett. B,* 34 (2020) 2030002.
24. Mamedov, B. A., Esmer. M. Y., On the Philosophical Nature of Einstein's Mass-Energy Equivalence Formula E= mc^ 2, Found. Sci. 19 (2014) 319-329.
25. Barish, B. C., Weiss, R., LIGO and the detection of gravitational waves. *Phys. Today*, *52* (1999) 44-50.
26. Scientific, L. I. G. O., Collaborations, V., Abbott, B. P., Abbott, R., Abbott, T. D., Abernathy, M. R., ... & Cepeda, C. B. (2016). Tests of general relativity with GW150914. Phys. Rev. Lett., 116(22) (2016) 221101.
27. Weiss, R., Gravitational radiation. *Reviews of Modern Physics*, *71*(2) (1999) S187.
28. Abramovici, et. al., LIGO: The laser interferometer gravitational-wave observatory. *Science*, *256*(5055) (1992) 325-333.
29. Braginsky, V. B., Gorodetsky, M. L., Khalili, F. Y., Matsko, A. B., Thorne, K. S., & Vyatchanin, S. P., Noise in gravitational-wave detectors and other classical-force measurements is not influenced by test-mass quantization. *Physical Review D*, *67*(8), (2003) 082001.
30. Einstein, A., Podolsky, B., Rosen, N., "Can Quantum Mechanical Description of Physical Reality Be Considered Complete?" Phys. Rev., 47 (1935)777.
31. Schrödinger, E., Discussion of probability relations between separated systems. Math. Proceed. Cambridge Philos. Soc., **31** (4) (1935) 555–563.
32. Bell, J.S., On the Einstein-Podolsky-Rosen Paradox, Physics,1 (1964)195.
33. Bell, J. S., On the problem of hidden variables in quantum mechanics. Rev.Mod. Phys., **38** (3) (1966) 447–452.
34. Schirber, M., Nobel Prize: Quantum Entanglement Unveiled. *Physics*, *15* (2022) 153.
35. Whitaker, A., The New Quantum Age, Oxford, 2012.
36. Freedman, S. J.; Clauser, J. F., Experimental test of local hidden-variable theories. Phys. Rev. Lett., 28 (938) (1972) 938–941
37. Clauser, J. F., Shimony, A., Bell's theorem: Experimental tests and implications,. Rep. Prog. Phys., **41** (12) (1978) 1881–1927.
38. Aspect, A., Dalibard, J., Roger, G., Experimental Test of Bell's Inequalities Using Time-Varying Analyzers, Phys. Rev. Lett., **49** (25) (1982) 1804–7.
39. Pan, J-W., Bouwmeester, D., Daniell, M., Weinfurter, H., Zeilinger, A., "Experimental test of quantum nonlocality in three-photon GHZ entanglement". Nature. **403** (6769) (2000) 515–519.
40. Dehlinger, D., Mitchell, M. W., Entangled photons, nonlocality, and Bell inequalities in the undergraduate laboratory. Amer. J. Phys., **70** (9) (2002) 903–910.
41. Gerhardt, I., Liu, Q., Lamas-Linares, A., Skaar, J., Scarani, V., et al. Experimentally faking the violation of Bell's inequalities. Phys. Rev. Lett., **107** (17) (2011).170404.
42. Aspect, A., Closing the Door on Einstein and Bohr's Quantum Debate. Physics. 8 (2015) 123.
43. Schlosshauer, M., Kofler, J., Zeilinger, A., A Snapshot of Foundational Attitudes Toward Quantum Mechanics. Stud. History and Philos. Sci. Part B: **44** (3) (2013)222–230.
44. Gröblacher, S., Paterek, T., Kaltenbaek, R., Brukner, Č., Żukowski, M., Aspelmeyer, M. Zeilinger, A., An experimental test of non-local realism. *Nature*. **446** (7138) (2007) 871–5.

45. Greenberger, D., Horne, M.A., Zeilinger, Z., Going beyond Bell's Theorem, I n Bell's Theorem, Quantum Theory, and Conceptions of the Universe, M. Kafatos, ed., Kluwer Academic, Dordrecht, The Netherlands, 1989.
46. Higgs, P.W., Broken symmetries and the masses of gauge bosons, Phys. Rev. Lett., 13 (1964) 508.
47. Englert, F., Brout, R., Broken symmetry and the mass of gauge vector mesons, Phys. Rev. Lett., 13 (1964) 321.
48. Atlas Collaboration, Aad, G., Abajyan, T., Abbott, B., Abdallah, J., Abdel Khalek, S., ... & Bangert, A., A particle consistent with the Higgs boson observed with the ATLAS detector at the Large Hadron Collider. *Science*, *338*(6114) (2012) 1576-1582.
49. Gagnon, P., Who Cares about Particle Physics?: Making Sense of the Higgs Boson, the Large Hadron Collider and CERN. Oxford University Press., 2016.
50. Fitch, V. L., Marlow, D. R., Dementi, M. A. *Critical problems in physics* (Vol. 34). Princeton University Press. 2021.
51. Troitsky, S. V., Unsolved problems in particle physics. Physics-Uspekhi, *55*(1) (2012) 72.
52. Ballesteros, G., Redondo, J., Ringwald, A., & Tamarit, C., Standard Model—axion—seesaw—Higgs portal inflation. Five problems of particle physics and cosmology solved in one stroke. *J. Cosmol. Astropart. Phys.*, *2017*(08) (2017) 001.
53. Hammond, R., The Unknown Universe: The Origin of the Universe, Quantum Gravity, Wormholes, and Other Things Science Still Can't Explain. Red Wheel/Weiser, (2008).
54. Lorentz, H. A., The Theory of Electrons, Dover, New York (1953).
55. Heitler, W., The Quantum Theory of Radiation, Oxford University Press, Oxford University, 1950.
56. Landau, L. D., Lifshitz, E. M., The Classical Theory of Fields, Pergamon, New York 1987.
57. Hooker, S., Webb C., Laser Physics, Oxford Press, 2010.
58. Meschede, D., Optics, Light and Lasers: The Practical Approach to Modern Aspects of Laser Physics, Wiley-VCH, 2017.
59. Sokolov, A. A., Kerimov, B. K., On the Damping Theory of Particle Scattering by a Fixed Center. *Soviet Physics Jetp-Ussr*, *4*(6) (1957) 921-922.
60. Sokolov, A. A., Kerimov, B. K., On The Scattering Theory Of Dirac Particles Considering The Attenuation. In *Doklady Akad. Nauk SSSR*, 105 (1955, Lomonosov Moscow State Univ..
61. Sokolov, A. A., Kerimov, B. K., & Guseinov, I. I. Damping theory study of elastic scattering of Dirac particles with account of polarization effects. Nucl. Phys., *5* (1958). 390-400.
62. Sokolov, A. A., Kerimov, B. K.,. Effect of Damping on Polarization of Dirac Particles in Scattering. Sov. J. Exper. Theor. Phys., *6* (1958) 639.
63. Guseinov, I. I., New complete orthonormal sets of exponential-type orbitals and their application to translation of Slater orbitals. Int. J. Quantum Chem., 90(1) (2002) 114-118.
64. Guseinov, I. I., New developments in quantum mechanics and applications, AIP Confer. Proceed., 899 (2007) 65.
65. Guseinov, I.I., New Complete Orthonormal Sets of Exponential-Type Orbitals in Standard Convention and Their Origin, Bull. Chem. Soc. Jpn., 85 (2012) 1306-1309.
66. Guseinov, I.I., Combined open shell Hartree–Fock theory of atomic and molecular systems constructed from noncharged scalar nuclear Fermi particles. II. Fermi particles with rest mass $m \neq 1$, Phys. Essays, 28(2015) 179-181.

67. Mamedov, B.A., Israfil I. Guseinov: A pioneer of the quantum theory of atomic, molecular, and nuclear systems, Int. J. Quantum Chem. 114 (2014) 361.
68. Görgün, N. S., Guseinov, I. I., Aydın, R., The use of quantum damping self-frictional theory in a study of hydrogen-like atomic energies and forces. *Results Phys.*, *11* (2018). 128-130.
69. Sahin, E., Ertürk, M., Ozdogan, T., & Orbay, M., Exponential type orbitals with hyperbolic cosine function basis sets for isoelectronic series of the atoms Be to Ne. *Zeitschrift für Naturforschung A*, *78*(1) ((2023) 1-8.
70. Guseinov, I. I., Mamedov, B. A. Studies of quantum self-frictional atomic potentials and nuclear attraction forces in standard convention. J. Molec. Struct., 1080 (2015) 24-29.
71. Guseinov, I. I., Mamedov, B. A., Use of quantum self-friction potentials and forces in standard convention for study of harmonic oscillator. Ind. J. Phys., 91 (2017) 371-376.
72. Çopuroğlu, E., Mamedov, B. A. Developments in molecular electronic structure evaluation based on the self-frictional field with Slater-type orbitals. *Ind. J. Phys.*, *93* (2019) 7-14.
73. Guseinov, I. I., Mamedov, B. A., Copuroglu, E., Application of quantum self-frictional nonperturbative theory for the study of atomic anharmonic oscillator potentials and their arbitrary derivatives. *Ind. J. Phys.*, *95* (2021) 405-410.
74. Coskun, M., Erturk, M., Double hyperbolic cosine basis sets for LCAO. Molec. Phys., 120 (2022) e2109527.
75. Guseinov, I.I., Erturk, M., Application of Combined Hartree-Fock-Roothaan Theory to Isoelectronic Series of Atoms Using Noninteger n-Generalized Exponential Type Orbitals, *MATCH Commun. Math. Comput. Chem.* 61 (2009) 603-613.
76. Guseinov, I.I., Sahin, E., Erturk, M., An improvement on $\psi^{(\alpha^*)}$ -exponential type orbitals for atoms in standard convention. *Molec. Phys.*, 112 (2014) 35-40.
77. Guseinov, I.I., Dirac equation for particles with arbitrary half-integral spin, *Philosoph. Mag.*, 91 (2011) 40634072.
78. Guseinov, I. I., Use of group theory and Clifford algebra in the study of generalized Dirac equation for particles with arbitrary spin. *arXiv preprint arXiv:0805.1856* (2008).
79. Guseinov, I. I., Combined open shell Hartree–Fock theory of atomic–molecular and nuclear systems, *J. Math. Chem.*, 42 (2007) 177-189.
80. Guseinov, I. I., Combined open shell Hartree–Fock theory of atomic and molecular systems constructed from noncharged scalar particles, *Phys. Essays*, 27 (2014) 351-355.
81. Condon, E.U., Shortley, G.H., Theory of atomic spectra. Cambridge University Press, Cambridge, 1970.
82. Gradsteyn, I.S., Ryzhik, I.M., Tables of integrals, sums, series and products. Academic Press, New York (1980).